# A Review on Security and Privacy of Internet of Medical Things

Mohan Krishna Kagita[1], Navod Thilakarathne[2], Thippa Reddy Gadekallu[3], and Praveen Kumar Reddy Maddikunta[4]

[1] School of Computing and Mathematics, Charles Sturt University, Melbourne, Australia.
[2] Department of ICT, University of Colombo, Sri Lanka.
[3,4] School of Information Technology and Engg, Vellore Institute of Technology, India
mohankrishna4k@gmail.com, navod.neranjan@ict.cmb.ac.lk,
thippareddy.g@vit.ac.in, praveenkumarreddy@vit.ac.in

**Abstract.** The Internet of Medical Things (IoMT) are increasing the accuracy, reliability, and the production capability of electronic devices by playing a very important part in the industry of healthcare. The available medical resources and services related to healthcare are working to get an interconnection with each other by the digital healthcare system by the contribution of the researchers. Sensors, wearable devices, medical devices, and clinical devices are all connected to form an ecosystem of the Internet of Medical Things. The different applications of healthcare are enabled by the Internet of Medical Things to reduce the healthcare costs, to attend the medical responses on time and it also helps in increasing the quality of the medical treatment. The healthcare industry is transformed by the Internet of Medical Things as it delivers targeted and personalized medical care and it also seamlessly enables the communication of medical data. Devices used in the medical field and their application are connected to the system of healthcare of Information technology with the help of the digital world.

**Keywords:** Internet of Medical Things, Internet of Things, Healthcare, Medical, Quality of Life.

## 1 Introduction

A network of medical devices and people are envisioned by the Internet of Medical Devices (IoMT) by which the data of healthcare are exchanged with the help of wireless communication [61]. Since the advanced technology and the population are increasing and growing all over the globe day by day the services of the healthcare sector and the cost of healthcare are also increasing every day[7]. In the coming days, if the healthcare sector and the internet of medical things work together then the quality of life can be improved, much better care services will be offered to the population and more systems can be created that will be cost-effective too[58].

---
*

The lives are changing in such a way that was never imagined by anyone in the past with the help of IoT and IoMT[14]. Not like the regular paradigm, each object is connected in the world of the Internet of Things and every object is considered as the smart object. "Internet of Things" is a network of things that are physical and virtual, and which is dynamic and self-configuring as well. Communication protocols that are inter-operable, media, and standards are the power backups of the Internet of Things. "Internet of things" is said to be the latest version of technologies such as information and communication[8]. The Internet of Things are having the identities and attributes that can connect the network of information like the internet, and able to work as sensors, data processors, networking, and communication tools. New challenges are opened in the domain of the healthcare sector by the Internet of Things (IoT), in terms of appliances that are smart, which is known as the Internet of Medical Things. Significant advantages are offered by the Internet of Medical Things for the people's well-being by improving the quality of their life and dropping down their medical expenses.

Since the number of patients are increasing every day, the regular and traditional systems of the healthcare sector are suffering so many challenges. The Internet of medical things are introduced so that the issue and challenges can be addressed, and the accuracy, reliability, and efficiency can be increased by equivalent time, the effectiveness of the domain of the health care sector can also be increased. To give a much more effective and efficient response to the needs of the patient's Internet of medical things are considered as the process of enhancement and investment as well[59]. There are various issues and the challenges that are faced by IoMT like no proper security system, problems in privacy measures, no proper training, and awareness regarding IoMT. Several applications of the Internet of Medical Things that are very closely associated with the sensitive ser- vices which are related to the healthcare sector are handling the information of the patients which are very private and sensitive such as the name of the patient, their address, and their conditions related to health[60]. To preserve the privacy of the patient without going down the level of security is the most challenging part of the domain of IoMT. It is very important to have a solution for security and privacy in IoMT which includes very little computation involvement, and which requires very few resources[58].

In the smart ecosystem of healthcare, the major step is the utilization of the potential of the technologies that already exist, to deliver the services that are best for the users and improve their life in a much better way. Wireless sensors are very critical elements. They are used to monitor the health status of the patients remotely and then communicate the information to the caregivers by the help of information and communication technologies[3]. The Internet of medical things acquires significant help from an empowering technology like artificial intelligence which is assisting the experts of the medical field in their proficiency areas like their clinical decisions[52][15]. By the use of the data which is generated by the professionals of the healthcare feedback the feed backs of the patients, the computers are learning the normal and abnormal decisions with the help of techniques such as machine learning and deep learning.

There is so much use of sensors of Body sensors in the Medical field to get the data of the patient easily from distant places instead of connecting with several biosensors and utilizing lots of money. Due to the issues of hacking, there is a threat of security while accessing the body sensor network, therefore, the sensitive details of the patients are secured by developing a secured fuzzy extractor and combining it with a fuzzy vault to provide more security by the use of bio metric key authentication system[26].

Since there is a huge development in the communication system in recent years, there is wide use of the internet of medical things in the healthcare sector. With the help of IoMT, the health data of the patient can be continuously monitored without the involvement of the laboratories by the use of wireless healthcare techniques and it will also improve the quality of life all over the world[52]. Body sensor network helps to monitor the health data of the elderly people and the children as well by their well-wishers in a much effective manner since they are considered as the major population of the globe and are the main concern by their well-wishers in terms of their health issues. IoMT is dependent on the healthcare systems that are using physical information like body temperature, skin resistance, signals of ECG, and EEG which is taken as the inputs by the sensor network known as Body Sensor Network or BSN[41].

## 2 RISKS OF INTERNET OF MEDICAL THINGS

There are so many risks that are associated with the deploy- meant of IoT of medical in the domain of healthcare. Some of them are:

1) The medical condition of the patient and the reputation of the hospital can be seriously affected if the personal information of the patient is disclosed.

2) The medical complication will be generated if the transmitted data is altered and modified by any medical device due to the falsification of data which will lead to the incorrect medical description and high drug dose[53].

3) Sometimes the employees that are unsatisfied with their job or when they are associated with some organized crime activity or being bribed leaks the medical details and the crucial information of the patients which leads to the risk of privacy and life of the patient.

4) Due to lack of training the nurses and doctors can bring in permanent disabilities or even sometimes losing the life of the patient.

5) The life of the patient can be seriously affected and sometimes they are led to disabilities and fatalities due to the inaccurate medical operations done by some specialized robots[12].

Despite all these challenges and risks, it is seen that the Internet of Medical Things is the combination of reliability and safety of the devices that are used in the medical field that are traditionally used and the dynamic, scalable and generous capability of traditional IoT. A combination of "Internet of medical things" and "Internet of Things" are solving issues of the elderly and the diseases that are chronic by managing so many devices that are deployed for many patients, and at the same time, they are enough generic to look after so many diseases which require hydrogenous monitoring and actuation. Standardization is the major challenge that is faced by the Internet of Medical Things. It is very important to have the standardization of all the medical devices that are working together, and the vendors need to go for the correct security measures to have the protection of getting hacked[20]. The Internet of medical things are also providing the solutions to the challenges like the mobility of the patient which means that if the patient is unable to move, even then he can be monitored, and medical facilities can be provided at their comfort zone by the use of IoMT. Instead of having so many issues, risks and challenges, there is a demand for the latest technological issues for the systems of healthcare, developing are trying to give the solutions to these visitors to change the way of delivering better healthcare to the world.

Since the use and development of medical devices that are connected and distributed, the advent of IoT of medical is caused and this also brings in the applications that are promising and potential but are challenging at the same time[11]. There is so much use of a network of Body sensors on the Internet of Medical Things to get the data of the patient easily from distant places instead of connecting with several biosensors and utilizing lots of money. The integration of devices that are wearable to the IoMT is more focused instead of medical devices that are personal since they are often coming as wearable devices. Sensors, wearable devices, medical devices, and clinical devices are all connected to form an ecosystem of the Internet of Medical Things. There are some concerns which need to address by the biomedical devices because of the strict ethical standards of the medical community. These concerns are

1) Reliability: Each time the functional goals need to achieve by a reliable system, which means that in normal operating circumstances unexpected failure should not accrue. It is mandatory to be reliable for the potential diagnostic nature of systems of IoT of medical so that it can give guaranteed information that is collected.

2) Safety: The operating system should not be harmed in any way by a safe system. In terms of the Internet of medical things, particularly for the medical actuators, it should have the proven facts that the users will not be harmed by the system in any case.

3) Security: The system related to the medical field should all must have protection from the threats and attacks from the external devices since they are collecting very sensitive and personal information of the users[57].

## 3    LITERATURE REVIEW

**Al-Turjman, Nawaz, and Ulusar** (2019) studied the status of the internet of medical things for the industry of healthcare; they had also studied the research and plans of development and their applications[2]. Instead of having so many technical and design issues and challenges, there is an exponential increase in the implementation of the Internet of medical things in the healthcare sector. Challenges are depicted and the generic framework of the "Internet of medical things" are shown which is composed of three major components to demonstrate the major challenges like the acquisition of data, gateways for communication, and the servers or the cloud[56]. Sensory technologies, wearable and non-wearable sensors are used to collect the data. To monitor a patient, it is good to use the wearable sensors since it will not interrupt the day to day activities of the patients. It is found that these technologies are helping caregivers and also dropping the cost of healthcare. A long and short-range of a communication network for the Internet of Medical Things in different cases is surveyed. It is found that WANs are the very reliable, trustworthy, and robust infrastructure of networks in the industry of healthcare when it comes to the accuracy of the data, mobility, and the safety of the patients but it is seen that they are consuming so much of power[42]. It is also found that the Decision Tools are a very good tool to make clinical decisions and enhance the services related to healthcare as it is effective in terms of cost and efficiency as well. Since there is no whole sole model of DM which can perform consistently for all the sets of data and therefore it is advised to the hybrids and combination of various tools that can support all the systems.

**Joyia et al.** (2017) found that there are so many people who had contributed to the Internet of medical things in the domain of healthcare and there are so many applications and challenges of IoMT as well when it comes to the services related to the medical and healthcare sector[33]. It is seen that the Internet of medical things has so much potential in the medical domain and the work will help the researchers the ones who are working in this field to identify the challenges on the Internet of Medical Things. The researchers will get to know about the applications of IoMT and its use in the healthcare sector.

**Kumar and Parthasarathy** (2020) studied the Body sensor network which has the scheme of secure authentication and effective as well [36]. The algorithm which is proposed by the researchers can solve the problems of damage of data, consumption of more energy, and the delay in the network in a much effective manner[24]. Discrete wavelet transforms are used to extract the features in the signals[16]. Users can decide the level the filters in the algorithm and the features extraction accuracy is increased when the level of filters is increased. The noise ratio of the signal is responsible for the efficiency of feature classification which can be removed by using the technique of effective adaptive filtering[22]. It is found that the noise is reduced to up to four percent by this filter in comparison to other techniques that are used previously that is why it is used to lower down the noise in input signals which are complex as a much effective tool. The techniques which are proposed in the study show true positive with an average of forty percent and a false positive rate which is also reduced in comparison to the Bayesian network which was already in use. The total energy consumption is also reduced by up to twenty percent[62][17]. The authentication phase and the key generation of the system are aimed at this method. A physical layer needs to be developed in the time ahead and aa more effective communication protocol will be required for the improvement of security in MAC and besides, to achieve the enhanced overhead efficiency and effective routing protocol will be required to be designed in the time ahead.

**Yaacoub et al.** (2019) concluded that there is a requirement of the design of intrusion detection or a prevention system which is more efficient, and which can cooperate with the honey spots having a dynamic shadow [40]. A security solution is proposed which has five layers for the detection and from the prevention of attacks and it should have the ability to reduce and correct the damage which is the result of these attacks and preserve the privacy of the patients at the same time[25]. But it is said that the major challenges which are still surrounding the Internet of Medical Things are zero-day attacks and exploits.

**Alsubaei et al.** (2019) found an Internet of Medical Things Security Assessment Framework which a web-based application and is dependent on the novel ontological environment and it ensures the security of the solutions on the Internet of Medical Things [1]. In detail, a list of attributes of assessment is recommended by the Internet of Medical Things security assessment framework in which all the essential security measures are covered. This allows the adopters of the Internet of Medical things to select and forces the security in solutions of the Internet of Medical Things which is based on their security aims and which is can be altered in a different scenario. The novelty of the Internet of Medical Things Security Assessment Framework is in their ability to adopt the latest technologies that are emerging and the new stakeholders, granularity, and standard compliance[44]. There are cases where the administrators of the system are the all alone responsible for the making the decisions that are related to the security, but an excellent opportunity id provided to the stakeholders by this work to had personal experience in the cut-throat sector of security on the Internet of Medical Things. Stakeholders are allowed to get to know about the associated risks which are there with the medical devices and they can follow best practices to overcome the risks and go for better decisions.

**Cecil et al.** (2018) studied that in the framework of cyber training there is a very limited scope of surgery training [5]. The professional is trained to do the surgical process in orthopedics only for plastic surgery which is majorly done for the fracture of the femur bone. GENI dependent networking technologies are playing a great role and support the activities of cyber training[9]. The design of the simulation environment is discussed in detail[23]. In a medical university, the residents of the orthopedic field are using the framework which is based on internet of medical things for their training activities; in their assessment part of studies, they are adopting the frameworks which are based on Internet of Medical Things (IoMT) in their surgical training and education.

**Elhoseny et al**. (2019) studied a group of experimentation which is done by the use of data which is collected from the women who are having a high risk of ovarian cancer due to the family history or history of cancer for an individual [27] [28]. While comparing the data with the recurrent neural networks which are also known as RSS with the feed-forward neural network (FFNN) and with some others, it is seen that the method which is proposed is attaining Amax accuracy of the sensitivity of 96.27 and 85.2 of the rates of specificity[62]. The results of the experiments show that the model which is proposed is very useful in detecting cancer at their initial stages. The model can detect the er with high accuracy, root means square with low error, sensitivity, and specificity as well.

**Farahat et al.** (2018) studied a model that has real-time security and it also has an encryption method which is encoded with the authentication [28]. This model can solve the security problems of Internet of Medical Smart Things (IoMST). Internet of Medical Smart Things is facing a very big issue of real-time security and privacy as well[50]. A technique of run-length encoding is presented in the study which is followed by the encryption that has a key that is rotating in the system of the patient and the data is decrepitude which has a rotated key and it is decoded with the decoding run-length technique in the system of the physician. The integrity of the record is ensured by the digital signature of the patient. Further, the study can be done by focusing on developing a shield of E-health which will collect all the data from the sensors of the Internet of Medical Smart Things and a system is developed by them who can do the encryption by using the asynchronous method. If the number of sensors is increased in our system, it can be implemented for solving the problems of big data. Then it is possible to implement a personal blockchain of the health record of the patient[21].

**Gatouillat et al.** (2018) revealed that the validation and verification of the system can also be done along with the robustness, security, and reliability of the system by an approach of cyber-physical systems CPS) [29]. While designing a biomedical system these are the crucial questions, cyber-physical systems are the much better process of designing for implementing, designing, and deploying such kind of systems in the biomedical field.

**Guan et al.** (2019) found that the technique of machine learning can be used for data processing which is involved in the Internet of Medical Things for the analysis of the medical data and diagnosis of the diseases [30]. At the same time, it is very important to pay great attention to the personal privacy information disclosure so that the personal and sensitive medical data of the patient and the hospital is not leaked out. The technique of cluster analysis plays an important role in e diagnosis of diseases and medical analysis. Efficient Def- eventually Private Data Clustering scheme (EDPDCS) which is based on the Map-Reduce framework is proposed in this study to allow the Privacy-preserving cluster analysis on the Internet of Medical Things. It is found that when the data sets of the algorithm of normalized Intra-cluster variance is compared with other data set of the algorithms it is seen that the accuracy of the deferentially private k- means is improved by Efficient Deferentially Private Data Clustering scheme (EDPDCS).

**Dimitrov** (2016) found that the new category of advisors of digital health will emerge in the future. This category of advisors of digital health care is skilled and they had they are capable of interpreting the data of the health and well-being of the patient [41]. They are helping their clients to avoid their illness which is chronic in nature chronic and their diet-related problems, cognitive function are improved, in achieving mental health, which is improved, and they are also helping their clients in achieving an overall improved lifestyle. Such roles of the Internet of Medical Things are becoming very important due to the elderly of the world population.

**Jin et al**. (2019) found that the prediction of the quantity of the patients from the outside is not a problem of simple time series and there are so many factors of different variety which are nonlinear, and they are influencing this problem[32]. At the same time in some other experiments, a prediction model was proposed which is multi-dimensional it has a feature of air quality which is much better than other models. Through the analysis of the data of the experiments, it is found that a very important and significant role is played by the indicators of air quality in the prediction of the respiratory clinic of the patients from outside and their visit to the clinic. The prediction of respiratory consultations and its significant lag effect and a lag prediction of four days which has a better effect are discussed in the paper. On the analysis of the results, it is seen that there is a lagging effect of air in respiratory diseases.

**Liu et al.** (2019) studied the involvement of Medical IoT of medical in the lung cancer treatment and the potential application of models of deep reinforcement learning in the diagnosis which involves the computers[34]. Models of deep reinforcement learning are presented in particular which had promised to be used in the localization and treatment of lung cancer. If the tumor of the lung is detected and diagnosed at their initial stages, then the effect of the treatment is improved and there is a chance of prolonging survival in much significant manner. The open challenging matters and applications of the model of deep reinforcement learning for the diagnosing and treating of lung cancer are discussed further in this study. It is seen that there are two major research directs thatch can be possibly proposed in the future which involves the localization of the tumors and the making the plan of treating it.

**Wang et al.** (2020) studied and designed an eencoderdecoder structure that is densely connected This structure shares the codes and at the same time integrates the decoders the depths which are different from each other[38]. The structure which is densely connected is correlated by the output of the multiple decoders [51]. An adaptive segmentation algorithm was proposed by the use of this structure for shallow and deep features. The advantages and benefits of the model which is proposed for the complex boundary segmentation were verified when the two sets of data were compared in the horizontal and vertical manner. The analysis of the experiments demonstrates that the algorithm of the depth feature of adaptive segmentation uses the information of different depths and the results of the segmentation which were generated by the decoders of different depth in an effective manner and the final results can be analyzed and by this the accuracy of image segmentation is also improved.

**Wasankar, Gulhane, and Gautam** (2017) studied the system which is known as K-Healthcare which uses four layers to work more closely and because of which the storing and processing is done much efficiently and along with that the valuable data is retrieved[39]. So many services are provided by this remotely like to prevent and diagnose the disease, assessment of risk, patient health can be monitored, users can be educated and treated by this system. The work which is going on K-healthcare focuses more on the development and deployment in real means[43]. The software can be designed, or an application of the smartphone can be developed in which the data can be directly obtained from the sensors and processed automatically.

**Rubi and Gondim** (2019) found a platform that prepares the data for OLAP application automatically with the help of automatic recording, thus the privacy of the patient is granted[37]. The application of data mining is simplified by the file system, the techniques of knowledge extraction, and the machine learning training of pre-processed automatic recording has resulted from the application of a repository of healthcare[10]. A framework of development was implemented for the devices of "Internet of Medical Things", gateway and the server for fog so that the new sensors of the platform can be integrated, an extension of functionalities of AE in the gateway and for the new binding protocols that can be implemented. Instead of isolated devices, the healthcare system is enabled which focuses on the data interchange.

**Islam et al** (2015) surveyed the healthcare technologies that are based on the "Internet of Things" and their diverse aspects and presented the different architectures of healthcare network along with the platforms which support the backbone of "Internet of Things" and facilitates the transmission and reception of medical data[31]. There are so many researches and development efforts that are done in the healthcare services and the applications that are driven by the "Internet of Things"[55]. The details of research activities that are concerned about the Internet of Things and its address to the pediatric and elderly care, supervision of chronic diseases, personal health, and fitness management is provided in the paper. There are so many works is going on to advance the sensors, healthcare devices, applications of the Internet, and other various technologies which will motivate the gadgets of healthcare that are affordable and expand the potential of the services that are based on the Internet of Things in the healthcare sector for their development in the future[54].

**Bhat et al.** (2017) address the necessity of the technology of "Internet of Things" with the solutions of e-Health and the devices that are wearable so the healthcare of the patients can be improved and for this, the access to EMR of the patient is provided which is quick and secure[4]. It is seen that the automation of building blocks of "Internet of Things" and the communication of Machine to the machine are continued as before but the layers of the services are added to complete the infrastructure[6]. A smarter approach for health-related services is not provided by the e-health system that is based on the "Internet of Things" but the process of decision making becomes more intelligent[18][19]. In all the aspects different issues of health are addressed by this system as a mass. As the model of e-health that is proposed has its foundation that is based on the digital world, the outputs are transformed to the second screen and the mobile devices in many easy ways[48].

**Polu** (2019) studied that the checking of the patient from the remote area is incredibly helping the patients and the healthcare experts as well but at the same time the accessibility of the RPM is yet not possible for the impacted individuals which can rely on the area they belong to and their abilities of remote access[35]. In addition, it is important for the experts to make much more effort with targeted goals to attract the patients and bring them in and spur them for utilizing RPM. At last, the precision of the gadget that is not proved yet is the fundamental disadvantage of this innovation. It is not known for how much time the imprecision will exist; the RPM will remain dubious for so many. The fundamental idea behind the framework that is proposed is to provide the statistics of health to the administrators of well-being and the patients are accurate and efficient after the execution of an organized cloud of data so that the information can be used by the experts and the patients and determine quick and effective action[47]. The final model which can be used by the doctors for the examination of the patients from anywhere on the globe at any time will be well equipped and loaded with all the features. In case of emergency, an emergency signal or a message will be sent to the doctors by this model along with the current status of the patient and medical information. For the easier and quicker access to this model from anywhere on the globe at any time, it can be deployed as a mobile app.

| No | Year/Author | IoT medical Sub Verticals | Technologies Used | Benefits of the Proposed System | Challenges in Current Approach | Solution for Current Issues | Drivers of IoT medical | Application |
|---|---|---|---|---|---|---|---|---|
| 1 | Alsubaei et al. (2019) | Security & Privacy | Web-based "Internet of medical Thing" Security Assessment Framework (IoMT SAF) | Security of stakeholders Decision making process | Security & privacy of the "Internet of medical Thing" Novel ontological scenario-based approach | Security features in "Internet of medical Thing" Protection & Deterrence in "Internet of medical Thing" | Adapt new Stakeholders & conformance to technology & medical standards. | Healthcare |
| 2 | Al-Turjman et al. (2019) | Patient monitoring, Sharing information | Sensory technologies, wearable & nonwearable Sensors. | Workload is decreased Decreased healthcare cost | High-power consumption. | A hybrid approach | Additional information is provided in healthcare information systems. | Healthcare |
| 3 | Banka et al. (2018) | Smart Health Monitoring | Raspberry pi | Recording of heart rate, BP of patients | | | Analyse &predict chronic disorders in the preliminary stage by the use of data mining techniques | Healthcare |
| 4 | Bhat et al. (2017) | Removes the physical limitations | Bluetooth Wi-Fi GUIs | Quick & secure access to patients' EMR | | Smart Approach | Transform the outputs to second screen & mobile devices | Healthcare |
| 5 | Cecil et al. (2018) | Orthopaedic Surgery | GENI-based Network | Surgical services by robots | Telemedicine approaches | | | Healthcare |
| 6 | Elhosenyet al. (2018) | Detection of Ovarian cancer | SOM ORNN | Early-stage detection of cancers with high accuracy, sensitivity | | | Classifies ovarian cancer data | Healthcare |
| 7 | Farahat et al. (2018) | Real-time Security & privacy | Run-length encoding technique | Solve the problem of the security of IOMST | Security Privacy standard Regulation | Security Privacy standard Regulation | Records the private blockchain of the patient's healthcare | Healthcare |
| 8 | Gatouillat et al. | Verification & Validation | CPS | Designing, implementing,test | | CPS Approach | Better control of | Healthcare |

| | | | | | | | |
|---|---|---|---|---|---|---|---|
| | (2018) | | | ing &deploying | | | system Robustness Security Reliability | |
| 9 | Guan et al. (2018) | Accuracy & Efficiency | Cluster Analysis | The accurate differentially private k-means algorithm | | A machine learning Approach | Accuracy & efficiency of clustering algorithm | Healthcare |
| 10 | Dimitrov (2016) | On-field mobile/tablet technology | Wearable & mobile apps | Hassle-free record management | Fiscal & policy issues | | Remote health monitoring | Healthcare |
| 11 | Islam et al. (2015) | Security & Privacy | IoT-based healthcare technologies | Paediatric & elderly care, chronic disease supervision, private health, & fitness Management | | | The intelligent collaborative Security model | Healthcare |
| 12 | Jin et al. (2019) | Analysis & Prediction | Intelligent sensing technology | Analysis & Prediction of medical data | Low quality of data acquisition | Use of BP neural network | Outpatient number prediction | Healthcare |
| 13 | Joyia et al. (2017) | | | Make life more convenient | Data privacy CPU capacity Security challenges | | | Healthcare |
| 14 | Kumar et al. (2019) | Network security | Body Sensor Network The effective adaptive filtering technique | Issues like data loss, energy consumption & delay in the network is solved effectively | Security threat | Fuzzy extractor combined with fuzzy vault approach | Access to patient's data remotely | Healthcare |
| 15 | Liu et al. (2019) | Lung cancer detection | Deep reinforcement learning models | Early detection & diagnosis of lung tumour | | | Lung cancer localization & treatment | Healthcare |
| 16 | Polu (2019) | Health monitoring | Wearable sensors | Remote monitoring | Unproved precision of gadgets | | To give accurate & efficient health statistics | Healthcare |
| 17 | Rubi et al. (2019) | Data analysis | M2M OLAP | Automatically prepares the data | | | Ensures interoperability, quality of the detection process | Healthcare |
| 18 | Wang et al. (2020) | Object detection | Encoder decoder structure UNet structure | Can extract various features from all layers adaptively | Medical image segmentation | Adaptive fully Dense (AFD) neural network for CT image segmentation | Accurate image segmentation | Healthcare |
| 19 | Wasankar et al. (2017) | Prevention & Diagnosis | RFID, IPv6, Cloud computing | Efficient storing, processing & retrieving of valuable data | | Data is directly Obtained from the sensors & processed automatically | | Healthcare |
| 20 | Yaacoub et al. (2019) | Safeguard & secure | | Train medical & IT staff | Privacy of patients, The confidentiality, integrity & availability of medical services. | Robust "Internet of medical Thing" | Secure & efficient system | Healthcare |

## 4 CHALLENGES FOR INTERNET OF MEDICAL THINGS

Standardization is the major challenge that is faced by IoMT. It is very important to have the standardization of all the medical devices that are working together, and the vendors need to go for the correct security measures to have the protection of getting hacked[49]. If all the devices are standardized then they will able to provide much more protected, efficient, scalable, consistent, and effective data. There are so many related but not limited challenges to the different constraints of the security of IoT of Medical Things. The Novel ontological scenario-based approach are the challenges of the present challenges of proposed systems. Fiscal and policy issues are most common challenges in the Internet of Medical Things. Low quality of data acquisition and high-power consummation, data privacy and CPU capacity security challenges and security threats, and unproved precision of internet of medical things and medical image segmentation are the main challenges of the present proposed systems[46].

## 5 PROPOSED SOLUTIONS FOR THE CURRENT APPROACH

Apart from all the success and advantages of proposed Internet of Medical Things technology is solving the present issues but with security and privacy concerns, there exist challenges aspect as well[46]. The main challenging issue includes the security issues as the main background to develop the approach based on cloud-connecting database. To overcome the security challenge, we have come up with a new approach that creates a dynamic, connected scheme to merge the advanced digital facilities allowing the Medical services. Moreover, further research should collaborate in creating a digital resolution like hybrid approach, telemedical approach, CPS approach, Machine Learning(ML), using BP neural acquisitions, fuzzy extractor combined with fuzzy vault approach, adaptive fully dense(AFD) can be used[13].

## 6 CONCLUSION

Internet of Medical Things is also known as healthcare IoT of medical which involves medical devices and their application that are connected to the systems of healthcare of Information technology with the help of the digital world. These devices and the systems are internet dependent which facilitates the communication between machines and for the storage of data they are linked to the digital platforms[45]. A wide range of clinical uses are involving Internet of medical Things such as smart watches and bracelets for medical alerts which can detect the falls of the patients and even, they can call the medical services in case of emergency. The healthcare industry is transformed by Internet Medical of Thing as it delivers the targeted and personalized medicine and it also seamlessly enables the communication of medical data. To monitor a patient, it is good to use the wearable sensors since it will not interrupt the day to day activities of the patient. It is found that the technologies that are involving Internet of Medical Things are helping the care givers and also lowering the cost of healthcare.